# A mechanism of vortex generation in a supersonic flow behind a gas-plasma interface

A. Markhotok

*Physics Department*
*Old Dominion University*
*Norfolk, VA 23529 USA*

## Abstract

The origin of a vortex structure generated during the shock-plasma interaction is investigated. A two-dimensional model based on the shock refraction mechanism successfully unifies the vortex generation with major co-processes typical for the interaction and thus well fits in their cause-and-consequence relationship. Numerical simulations demonstrated the possibility of an intense vortex generation with a continuous positive dynamics as the shock crosses the interface. It was shown that while the vorticity is triggered by the shock refraction on the interface, it is the gas parameter distribution that distinctively determines the parameters of the vortex evolution. The proposed model also provides an insight into interesting aspects of the refraction effects for both, the shock wave and the flow behind it (double refraction). The results are applicable to the problems of energy deposition in a hypersonic flow, a flame-shock interaction, in combustion, in astrophysics, and in the fusion research.

## I. INTRODUCTION

It is well known that a shock wave interaction with plasma results in significant changes in the shock wave structure and the plasma flow. Such phenomena are particularly observed in the shock–flame interactions [1], in the front separation regions control experiments [2], in combustion [3,4], and in the electric discharge [5], RF- [6], or laser-induced [7,8] energy deposition experiments. Dynamic instability and turbulence in impulsively loaded flows is also of considerable interest in astrophysics plasmas [9] and fusion research [10,11].

The complex nature and a number of co-processes often involved in this type of the interaction can be seen from the images of the initially simply structured planar, bow, or oblique shock evolving into a complicated system of distorted and secondary shocks with flow separation regions and formation of vortices [12]. Among the most remarkable changes are: the shock wave acceleration and its strong front distortion  increasing with time followed with remarkable weakening of the shock until it appears less and less identifiable [12-15]; motion of the shock away from the body in the presence of heating [16]; substantial changes in the gas/plasma parameter distribution behind the shock, particularly sharp reduction of pressure [17]; remarkable, up to 40% reduction in the wave drag experienced by a body when the plasma is created upstream [18,19]; the time-delays in the effects on the flow relative to the discharge on-off times and a finite pressure rise time [7]; and a vortex system formation in the flow behind the shock often followed with strong distortion or collapse of the plasma region [4,7].

For many experimental conditions, the thermal heating creating the plasma region can be the main cause affecting the changes in the shock structure and the flow parameters. The causes for the changes discussed in the current literature were the Mach number decrease, possibility of mean molecular weight and number density due to molecular dissociation and ionization, significant pressure drop attributed to the generation of vortex formed during the interaction of the thermal spot with the body shock [20], generation of a shock wave by a discharge and formation of a heated channel behind the energy source that interacts and diverts the body shock [17]; formation of the flow separation region [17]; and formation of a subsonic tube of a finite size with the supersonic flow outside of the tube and the shocks forming away from the heating zone [16].

A possible non-thermal nature of the interaction has been discussed for flows involving atomic and molecular transitions, gas kinetics, electrical properties of plasmas, and non-equilibrium states as a result of fast evolving processes such as radiation or fast expansion. Among them are: appearance of charged particles leading to upstream momentum transfer in the hypersonic flow [17]; the possibility of deflection of the incoming flow by plasma in front of the shock via electronic momentum transfer collisions [17]; and the release of heat into the shock layer by the exothermic reactions increasing the shock layer temperature and thus reducing the pressure and the density behind the shock wave [21]. The relationship between the shock wave refraction and all the changes in the shock structure and the flow has been discussed in [21,22]. It was shown that a relative curvature between the shock front surface and the interface, along with a steep density gradient across the interface, can be responsible for the chain of subsequent transformations in the flow, such as typical deformations of the front seen as the odd shapes of the deflection signals, the shock wave weakening/extinction in the plasma area, possibility of its restoration at the exit, and changes in the gas pressure and density in front of the body leading to the wave drag reduction [22].

A vortex system in the form of intense toroidal vortex rings developing in the plasma region is often observed simultaneously with all the processes accompanying the shock-plasma interaction [2,7]. Those complex phenomena became very important because of their presence in a class of problems involving the shock waves propagation through hot gas or plasma. The instability induced by the shock wave passage through a flame is one of the basic phenomena applicable to the supersonic combustion. Shock-flame interaction provides a means for the transition to turbulent flame in combustion systems, allowing for the increased chemical reaction rates. A vortex generated by a passage of the shock leading to transition into a turbulent flame is known to result in a dramatic increase of the burning velocity [23,24]. The effect of the strength of the shocks passing through the flame zone on the total burning velocity increase was studied computationally in a supersonic combustion problem involving multiple flow deformations [24]. The authors noted the possibility of the effects of flame distortion and reflecting shock waves on the total circulation in the flow, the shock strength influence on the vortex generation and flame distortion, and a possible relation between the local burning velocity of the distorted flame and the laminar burning velocity.

A vortex-related turbulence that could enhance local combustion rates was studied in [25,26]. The gas-dynamic distortions were found having obvious implications in chemically reacting systems. The initial experiments showed that accelerated reaction rates are linked to local vorticity and reactivity of the initial mixtures. For reactive mixtures, such as acetylene-air, the increases in the combustion rates are found to be due to vorticity "burned-out" in the toroidal mixture and re-establishing a more spherical flame bubble. In lower reactivity mixtures, such as methane-air, the macroscopical toroid form can be retained and an attempt to quantify the combustion rate enhancement in terms of the local increase in vorticity was made. An increase in chemical reaction rates and an important relationship between burning velocity and vorticity/turbulence have been also confirmed in the studies [27,28].

Among the mechanisms that can be responsible for the vortex production determined by conditions on an interface is the baroclinic effect that is due to non-alignment of pressure and density gradients in the plasma region. Sharp density gradients and the presence of ambient pressure/shock waves, along with a typical geometry (planar shock/pressure wave and cylindrical laminar flame fronts) make the baroclinic effect one of those often used mechanisms for the vortex production. The general vorticity equation for a compressible, viscous flow with variable fluid properties can be written using the total time derivative operator *D/Dt*:

$$\frac{D\bar{\omega}}{Dt} = (\bar{\omega} \cdot \nabla)\bar{u} - \bar{\omega}(\nabla \cdot \bar{u}) + \frac{\nabla\rho \times \nabla p}{\rho^2} + \frac{1}{\text{Re}}\left(\nabla \times \left(\frac{1}{\rho}\nabla \cdot \vartheta\right)\right) + \Delta \times \bar{F} \qquad (1)$$

where $\omega$ is the vorticity, $u$ is the flow velocity, $\rho$ and $p$ are the local density and pressure, Re is the Reynolds number, and $\vartheta$ is the viscous stress tensor. The first term on the right hand side of the equation is due to vortices stretching/tilting, second is the bulk dilatation, the third is the baroclinic term, fourth is the viscous term, and $F$ is the sum of external forces. As seen from the equation, the baroclinic vorticity is the only source term responsible for triggering the vorticity in the flow. In the experimental studies of the vorticity induced in the axisymmetric flow by laser driven optical breakdown in quiescent gas [7], the shock incoming on the plasma sphere in the upward direction yields an effective negative pressure gradient in the *y*-direction, and the baroclinic term $d\omega/dt = (\nabla\rho \times \nabla p)/\rho^2$ has an effective negative component in the left half and positive in the right half of the sphere. This positive and negative vorticity was observed near the leading and trailing edges of the plasma with the magnitude of vortex observed to decrease in time as it spreads out further from the core of the plasma.

The effect of baroclinity in the spherical/cylindrical geometry of the flame front was also studied in [1] where the vorticity field rapidly distorted the laminar flame front and the induced vorticity eventually lead to the turbulent break up of a laminar flame. Similar interaction of a flat shock wave with a cylindrical flame/hot region was studied in [29,30]. A relationship between the magnitude of the total circulation due to the baroclinic effect $d\Gamma/dt = \int_A [(\nabla\rho \times \nabla p)/\rho^2]dA$ and the flow parameters were obtained for chemically nonreactive flows, with stoichiometric $H_2$/air mixture intended for possible fuel for supersonic combustion, and $\Gamma = 2U_1 R(1 - U_1/aMc_1)\ln(\rho_1/\rho_b)$, where $R$ and $\rho_b$ are radius and bubble/flame density, $\rho_1$ is the density of the pre-shock mixture, $U_1$ is the flow velocity, and $M$ is incident Mach number. Neglecting the boundary distortion and assuming homogeneous density distribution in the flame, the effect of shock strength on the flow was studied. It was found that for three different shock Mach numbers between 1.05 and 1.5, an increase of the shock strength greatly accelerates the flame distortion and the onset of the break-up into two separate vortices due to a significant intrusion of cold gas into the flame. In this case both, the total burning velocity and the length of the flame front increase dramatically with the shock Mach number.

The vortex structure formation as a result of shock-plasma interaction is also commonly attributed to Rithchmer-Meshkov instability (RMI) that occurs when two fluids of different density are accelerated by a passage of a shock wave. The development of instability begins with small amplitude perturbations of the plasma interface which initially grow linearly with time [8]. The acceleration of both gasses by the shock wave causes grows of initial periodic perturbations, which begin to increase in amplitude. This is followed by a nonlinear regime with bubbles appearing in the case of light fluid penetrating a heavy fluid, and with spikes appearing in the case of a heavy fluid moving into the light one. Finally, the vortex sheet rolls up and accumulates into periodic vortex cores in the post-shock flow. In experiments [7] the vortex generation in the plasma-shock interaction is thought to be a result of Richmeyer-Meshkov instability and turbulent mixing that was induced by baroclinically driven flow motion at later times of the interaction. Atwood and shock Mach numbers, as well as the laser energy-ambient pressure ratio, were found as primary control parameters. In the experiments, higher ambient pressure resulted in more small-scale perturbations, with the cause of this pressure effect not fully understood.

The two common mechanisms described above were used to successfully explain vortex formations in many experiments. At the same time, the full picture of the shock-plasma interaction tends to be more complicated, with a number of co-processes accompanying the vortex development [29-31]. Such a vortex system formation during a blunt body shock wave-plasma interaction has been studied in the experiment [29]. After a laser pulse was irradiated, the laser heated gas expanded through optical breakdown forming a close to spherical low density region upstream of the bow shock. Due to convection, the region collided with the bow shock with its shape transformed during their interaction. When plasma completely transmitted the bow shock wave, at about 92 μs, vortex ring of a donut shape was formed, and its size grew downstream, up to the model diameter size. Within the same times and simultaneously with the vortex formation, stagnation pressure significantly dropped to its minimum levels. The shock-thermal spot interaction causing a "lensing" of a blunt body shock simultaneously with a vortex pair that finally impinged on the blunt body surface was also observed in [32]. Upon passage of the vortex pair, a greatly distorted shock wave formed and propagated upstream.

The stagnation pressure drop in front of the body and decrease in the drag, in addition to the vorticity, is typically registered, as, for instance, in experiment [9]. In the study of the drag reduction effect of the blunt body by a single pulse energy deposition, the inviscid flow computation was performed for the interaction between spherical thermal spot and a bow shock generated in a supersonic flow field around a blunt body [33]. It was found that there was a strong relationship between vortex generated via unsteady hydrodynamic phenomena and drag reduction, and the amount of reduced energy considerably exceeded the deposited energy. The drag was reduced more when vortex structure was larger and these two phenomena were observed within the same time interval. Similarly strong correlation between vortex and reduced energy was observed in [33] where the vortex energy was proportional to the radius of the low density region and the free stream Mach number.

In experiments with a flame-shock interaction [30], a vortex system was also observed, and the flame distortion and the appearance of a secondary shock wave greatly affected the total circulation. After a shock wave passage through a flame, the flame was distorted and the shock was no longer planar, with a set of weak secondary shock waves being present behind the incident shock. The peak vorticity was located at the top and bottom of the cylindrical flame, and the vortexes were symmetrically rotating in opposite directions. The circulation changed its direction when the shock exited the flame bubble and a reflected shock occurred. The total circulation in the experiments was found less than expected due to flame distortion and the secondary shock waves.

The above observations show that the vortex production is not an isolated phenomenon. It is accompanied by a number of co-processes simultaneously present in the flow volume such as shock wave acceleration, its distortion and subsequent weakening, and secondary shock structures. This can hardly be explained with the baroclinicity or RMI, suggesting that they may be not the only causes of the vortex generation. Changes in the flow volume and remarkable dynamics in the vortex strength as the shock propagates well beyond the interface, suggest that there should be other, probably volume mechanisms of the vortex production continuously supplying it with the energy necessary for a positive dynamics.

Due to the coexistence of the vortex generation with other features of the shock-plasma interaction, a single mechanism relating the vortex generation to all those co-processes is thought. An attempt to connect a number of processes accompanying the shock–plasma interaction with the wave drag reduction has been already made [22].

In the present paper, a further attempt to relate those processes to the simultaneous vortex structure generation in the flow past the gas-plasma interface will be presented. It will be shown that the shock wave refraction on the interface can be the sole mechanism responsible for a vortex structure generation in the form of a toroidal vortex ring or two symmetrical vortex tubes rotating in opposite direction. The developed model based on this hypothesis will show both, the possibility for the shock refraction to trigger a vortex structure, and to unite the vortex generation with other co-processes accompanying the shock-plasma interaction. To separate the action of the "interface" and the "volume" factors, the model was developed using the same interface for three different types of the plasma parameter distribution. In this way, any significant difference in the results can be attributed to the volume, and any common features to the interface effects.

## II. THE MODEL OF VORTEX GENERATION

The model described here is designed to adequately describe the formation of a vortex structure following the shock wave passage through a gas-plasma interface and its further propagation through the plasma volume. It will be based on the assumption that all the changes in the shock structure and the flow are due to the shock wave refraction on the interface with plasma having a specified density distribution in its volume. The model will also show that the vortex system formation is in direct connection with other co-processes accompanying the shock-plasma interaction.

Because the discharge types used in the experiments typically produce a thermal spot of almost spherical shape [15,34,35], the derivations are done for a spherical geometry and the shock wave will be approximated also as spherical. The problem is considered in the vertical plane of symmetry so the relations are also valid for cylindrical geometry. It is assumed that the plasma cloud has been created in a discharge that is distant from the shock wave and it is moving with the flow toward the shock. When the cloud arrives at the shock location, the interaction between them occurs. It starts when the shock front first touches the plasma cloud boundary, and this is the moment when the time $t$ in the relations starts to be counted.

As shown in the Fig. 1, a spherical interface separates a hot plasma inside the cloud (medium 2) and a surrounding cold gas (medium 1). Both media are treated as ideal gasses with initially equal pressures on both sides of the interface. The temperature of the cold gas $T_1$ is assumed to be distributed homogeneously. $T_2$ is the hot plasma temperature right behind the interface that, in general, varies with distance from the interface following a definite law of parameter distribution. The temperature $T_2$ is higher than $T_1$, and its change across the interface can be abrupt (step–wise) or smooth [35]. The radius of the plasma boundary (orange curve in the figure) is denoted as $R_b$ and the radius of the incident shock front (black) is $R_s$.

In the reference frame stationary for the cloud, the spherical shock wave is incident on the spherical interface from left to right, center to center, with constant velocity $V_1$, as shown in the Fig. 1. After the interaction on the interface, a refracted shock wave (green curve) accelerates in the plasma medium to the velocity $V_2(x)$ that can be constant or dependent on time, depending on the plasma parameter distribution in the cloud [14]. For simplicity, the fact that the plasma sphere is continuously expanding will be neglected here and thus considered of a fixed diameter.

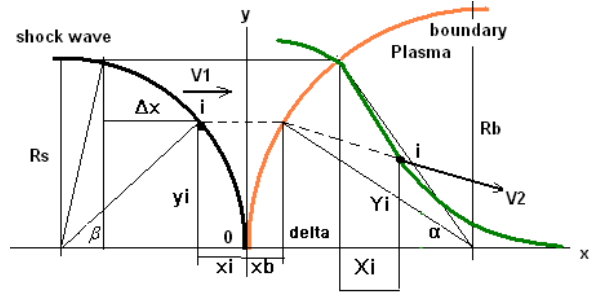

*Fig.1. Shock wave-plasma cloud interaction diagram in the vertical plane of symmetry. As the initially spherical shock progresses through the spherical interface (from left to right), its front shape gradually deforms (green curve) as a result of the interaction. Because of the symmetry, only the upper half of the diagram is shown.*

Consequently, these studies are applicable to a relatively slow developing gas/plasma cloud, or at later times of the cloud evolution when the expansion slows down and the system had enough time to achieve a thermal equilibrium state. This could be the case when thermal energy is deposited far from the body, the situation often reported by experimenters.

When a shock wave crosses an interface, its velocity vector changes its magnitude and direction (shock wave refraction) [35]. The turn of the shock velocity vector at an angle $\gamma_i$ varies with the location of the point of interaction $i$. Thus, even though the shock started its motion horizontally, after its refraction on the interface its velocity acquires a location dependent $y$-component. Then the upstream flow (behind the shock front) will also acquire two velocity components $v_x$ and $v_y$ that can result in non-zero vorticity

$$\omega = \frac{\partial v_x}{\partial y} - \frac{\partial v_y}{\partial x} \qquad (2)$$

To determine the velocity components $v_x$ and $v_y$, a transition from a laboratory reference frame used in previous calculations [35] to the frame moving with the shock can be made. In this reference frame, the velocity $V_2$ becomes the up-stream flow velocity and a known relation for the velocity of the flow behind a normal shock wave $v_n$ can be used

$$v_n = \frac{2}{k+1}\left(V_{2n} - \frac{a_c^2}{V_{2n}}\right) \qquad (3)$$

where $a_c$ is speed of sound. While the normal to the front component of the up-stream velocity $v_n$ undergoes changes across the front in accordance with the relation (3), its tangential component will be continuous, $\vec{v}_t = \vec{V}_{2t}$ (Fig. 2).

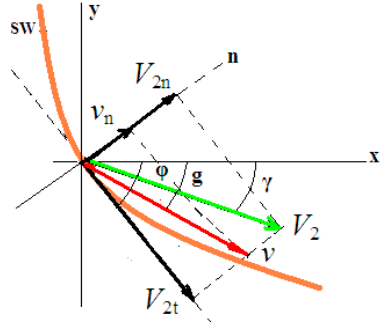

*Fig. 2. The refracted shock wave (orange curve) and up-stream flow velocities vectors diagram at the time when the shock propagates through the plasma cloud. Initially spherical front has been modified (stretched), as shown in the figure (the orange curve). Angles γ, φ, and g are the refraction, tangential to the front surface, and the up-stream flow rotation angles correspondingly.*

Using the geometry depicted in Fig. 2, the expressions for *x*- and *y*-component of the flow velocity behind the shock can be determined as

$$v_x = v_n \sin\varphi + V_2 \cos(\phi - \gamma)\cdot\cos\varphi$$
$$v_y = v_n \cos\varphi - V_2 \cos(\phi - \gamma)\cdot\sin\varphi \quad (4)$$

Here, φ is the angle between the tangential line to the shock front at a point *i* and the *x*-direction (the local front inclination angle), and γ is the refraction angle [36].

It follows from the equations that there is a turn for the flow velocity relative to its initial propagation direction at the angle

$$g = \tan^{-1}(v_y/v_x) \quad (5)$$

which occurs after the initial turn of the shock velocity vector (Fig. 2). Thus, in addition to the rotation of the shock velocity vector at an angle γ that happens on the interface (refraction), the second rotation at the angle *g*, now for the flow behind the shock, happens in the plasma volume. This second turn occurs across the shock front and is measured relative to the *x*-direction. Then the shock wave and the flow behind it will diverge at a relative angle ξ (double refraction):

$$\xi = \tan^{-1}(v_y/v_x) - \gamma \quad (6)$$

The phenomenon of double refraction, or two consecutive changes in the direction of hypersonic flow motion – first for the shock velocity relative to the incident flow direction happening on the interface, and then for the flow behind the shock - relative to the refracted shock velocity direction happening across the shock front, results in a rotational motion in the flow and gives rise to a non-zero circulation in the plasma volume.

To track possible space and temporal variations of the vorticity in the flow as the shock progresses through the plasma sphere, it can be calculated at specific locations along the shock front surface, at different interaction times. The shock front distortion during the shock-plasma interaction has already been studied in [22]. Then the vorticity produced at a specific point *i*, at a location of the shock front, can be calculated as

$$\omega_i = \omega(X_i, Y_i) = \left(\frac{\partial v_x}{\partial y} - \frac{\partial v_y}{\partial x_i}\right)_{x=X_i, y=Y_i} \quad (7)$$

where $X_i$ and $Y_i$ are the front coordinates and the angles $\gamma_i$ and $\varphi_i$ correspond to the same interaction point $i$. For the same "sphere-to-sphere" geometry considered in [22], the equations can be set for an interaction time $t = nR_b/V_1$, $0 < n < 2$, scaled with a characteristic time $\tau = R_b/V_1$

$$X_i = ((V_2/V_1)\cdot\cos\gamma - 1)(nR_b - (x_i + x_b)) + \Delta x \qquad (8)$$
$$Y_i = y_i - (V_2/V_1)\cdot\sin\gamma\cdot(nR_b - (x_i + x_b))$$

where the distances $\Delta x$ and $x_b$ (Fig.1) are determined as

$$x_b = R_b(1-\cos\alpha)$$
$$\Delta x = R_s(\cos\beta - \chi), \quad \chi = \frac{2(R_s + R_b)(R_s - nR_b) + n^2 R_b^2}{2R_s[(R_s + R_b) - nR_b]} \qquad (9)$$

The shock wave velocity and the refraction angle are determined by the problem geometry and heating intensity (across the interface) and account for the shock reflections off the interface through the ratio of Mach numbers in the two media [21]:

$$V_2/V_1 = \sqrt{\cos^2\alpha\cdot(T_2/T_1)\cdot(M_{2n}/M_{1n})^2 + \sin^2\alpha} \qquad (10)$$

$$\gamma = \alpha - \tan^{-1}\left(\sqrt{T_2/T_1}\cdot(M_{1n}/M_{2n})\cdot\tan\alpha\right) \qquad (11)$$

Here, $(x_i, y_i)$ and $(X_i, Y_i)$ are coordinates of the incident and refracted shock front surface at a point of interaction $i$, and $\alpha$ is the incidence angle at this point (Fig.1). It will be further assumed in the calculations that the interface is "smooth", the case when the refraction effects become more pronounced [37]. The media on both sides of the interface are considered as ideal gases with initially equal pressures across the interface.

At the second stage of the interaction, when the refracted shock wave starts to propagate off the interface through the plasma volume, the effect is dependent on the density distribution in the cloud [15,22,35] and will be considered separately for each type of the distribution. Even though the changes in the shock front shape start to appear during this period of time, they are still the consequences of the interaction at both stages, on the interface and in volume. To connect both stages of the interaction, the shock/flow parameters in the equations are tailored at the moment of crossing the interface, the same way as it was done in [22].

### III. NUMERICAL RESULTS AND DISCUSSION

The purpose of work in this section is to numerically verify if the shock wave refraction on the interface and further shock's interaction with plasma volume can produce a significant vorticity, of the size and the intensity level reported in the experiments. To split two factors of influence, the interface and the volume effects, the problem geometry, incident shock intensity, and the heating strength ($T_2/T_1$) will be kept fixed and the plasma density gradient will vary. The model will be run using three density distribution types covering common ways of the energy deposition/plasma creation. For drawing more general conclusions and broader comparison with experiments the geometry and the shock/plasma parameters will be kept the same as for results obtained in [22].

### III. A. The distribution of plasma parameters is homogeneous

Assuming the homogeneous gas parameter distribution on both sides of the "smooth" interface, the system of equations (8-11) can be employed to determine the flow velocity appearing in the equations (3-4). To recast it in a dimensionless form, the coordinates can be scaled with the radius $R_b$, the shock velocity with $V_1$, gas temperature with $T_1$, Mach number with $M_1$, and time - with the characteristic time $\tau = R_b/V_1$. Then the system (8-11) transforms into

$$\overline{X}_i = (\overline{v}\cos\gamma - 1)(n - (\overline{x}_i + \overline{x}_b)) + \Delta\overline{x}$$
$$\overline{Y}_i = \overline{y}_i - \overline{v}\sin\gamma \cdot (n - (\overline{x}_i + \overline{x}_b))$$
(12)

where $\overline{x}_b = 1 - \cos\alpha$, $\Delta\overline{x} = (R_s/R_b)(\cos\beta - \chi)$, and $\overline{V} = \sqrt{T \cdot M \cos^2\alpha + \sin^2\alpha}$, $n = t/\tau$. The vorticity can be scaled with the ratio $V_1/R$ resulting in its dimensionless equivalent $\overline{\omega} = \omega/(V_1/R_b)$.

Numerical results for the vorticity versus the coordinate $Y_i$ (7) presented in Fig. 3 were generated for $M_1 = 1.9$, $T_1 = 293$ K, $T_1/T_2 = 0.10$, $R_s = R_b = 0.3$ cm, and the adiabatic index $k = 1.4$ (air). Due to the symmetry, only the upper half of the picture is shown in the graph. Thus the results on the graph correspond to a vortex sheet with rotations in both halves of the picture in opposite directions (in cylindrical geometry) or a toroidal vortex ring if considered in the spherical geometry. The curves correspond to a set of propagation times starting at $n = 0.18$ (the most left curve) through 0.42, through the equal time intervals $\Delta n = 0.03$, and the last curve corresponds to $n = 0.44$ for better resolution on the graph.

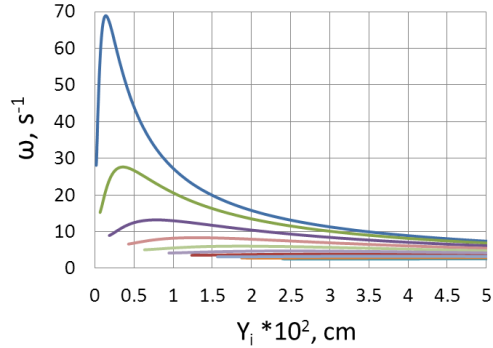

*Fig.3. The vorticity $\omega$ vs vertical coordinate Yi zoomed into a smaller region of the coordinate near the symmetry axis. The curves are obtained for different propagation times starting at n = 0.18 through n = 0.42, with $\Delta n$ = 0.03 increments, and the last curve corresponds to n = 0.44. The time sequence is from lower curve to upper.*

As seen from the figure, the spike in the vorticity intensity is observed in the close proximity to the symmetry axis, exactly where the abrupt change in the shock front structure occurs (Fig. 4a). This change in the front surface shape determined with the front inclination angle φ (Fig. 4 b) is the key factor pointing at the origin of the vorticity. The vortex intensity non-linearly increases with time (Fig. 3) revealing the possibility of a strong positive dynamics in its development.

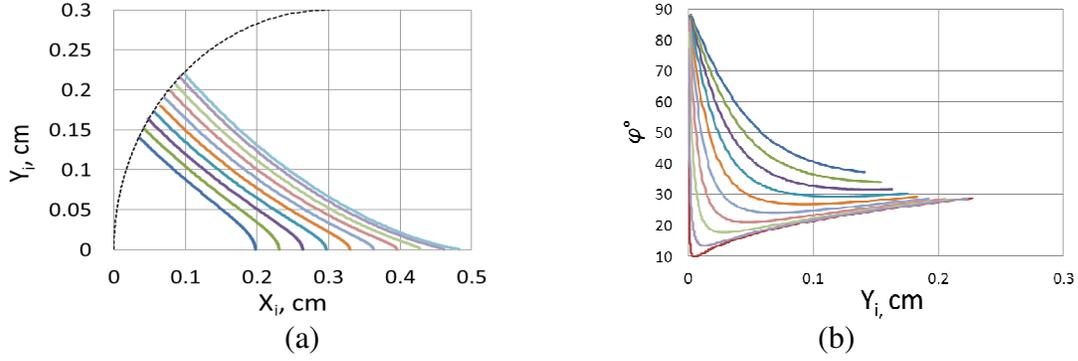

Fig. 4. (a) *Shock front transformation as the consequence of its refraction on the interface, for the same parameters as in Fig.3. A part of the shock front outside of the plasma sphere remains spherical (not shown in the picture) while the inside part undergoes strong deformation. The time sequence is from left curve to right. (b) The shock front inclination angles φ vs $Y_i$ corresponding to the curves in the graph (a). The time sequence is from upper curve to lower.*

Thus two key parameters are identifiable. The first one is the gradient of the angle φ distribution along the coordinate *Y* quickly increasing with time (Fig.4 b). The second parameter, location of the gradient maximum, exhibits a shift toward the symmetry axis (small *Y*'s) with time, the same trend as for vorticity. Thus the vorticity starts at an intermediate location in plasma cloud where there is a sharp distortion in the shock front, and then moves with time toward the symmetry axis, with its sharply increasing size and intensity.

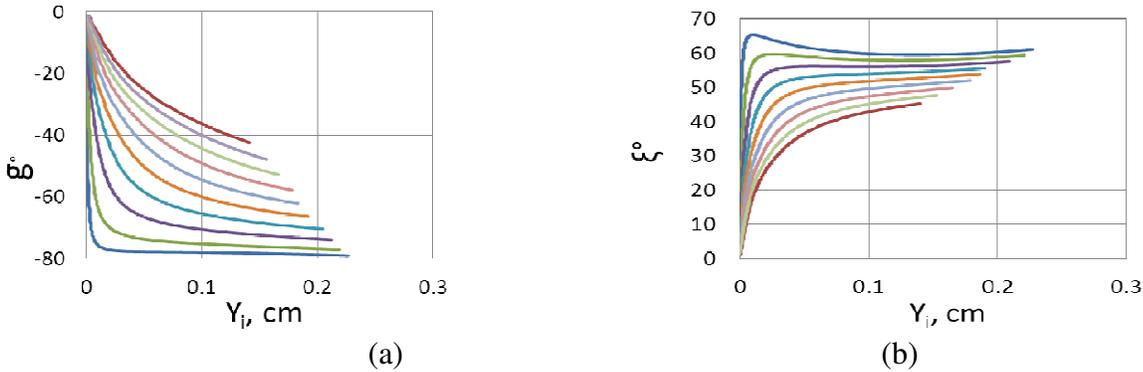

*Fig. 5. (a) The angle "g" determining the up-stream flow velocity direction, plotted vs the coordinate $Y_i$, for the same parameters as in Figs.3,4. The time sequence is from upper curve to lower. (b) The angle "ξ" demonstrating the divergence in the motion directions between the shock wave and the flow behind it. The time sequence is from lower curve to upper.*

An interesting effect of the upstream flow velocity vector rotation at the angle *g* defined in (5) is presented in the Fig. 5. As seen from the graph (a), the sharp increase in the rotation is happening at the same locations where the sharp changes in the shock front inclination angle φ occur (Fig. 4). This means that the vorticity develops due to significant flow rotation that occurs at the locations where there is a sharp distortion on the front.

The divergence angle $ξ = g − γ$ (Fig. 5b) displays the same pattern in its distribution and dynamics as the vorticity, with the intensity spikes that occur at the locations of the most intense distortions on the front (Fig. 3a). As seen in the figure, the amount of rotation is considerably higher, up to tens of degrees. It rises very sharply near the symmetry axis and develops at a short time scale (of a fraction of the characteristic time $\tau$).

As shown in [22], for this type of the distribution and problem parameters, the pressure in the flow behind the shock significantly drops and the wave drag is reduced proportionally. It is important that these changes are happening simultaneously with the vorticity development and are consequences of the same deformations on the front due to the shock refraction. This result identifies the same origin of the vorticity generation and a number of main co-processes observed during the shock-plasma interaction.

The remarkably strong volume effect found in the uniform plasma suggests the possibility of a non-zero gradient in the plasma density distribution to influence its strength. In the following two paragraphs, the vorticity generation in plasma with two types of non-uniform parameter distribution (exponential and power law) will be explored.

### III. B.  The exponential density distribution case

In the case of a non-uniform parameter distribution, the shock wave velocity becomes time dependent and the system of equations (8), (10), and (12) must be substantially modified depending on the type of the distribution [15,37]. Shock refraction effects on the interface are shown to be quite specific if the exponential plasma density profile is present [15]. Such a density distribution in a plasma can be established during the exothermal expansion and is often observed in large-area plasma sources created with internal low-inductance antenna units [6], detonation [3], or the ultra-intense laser-induced breakdown in a gas [38]. In the derivations below, the plasma density in the thermal spot is assumed to be exponentially decreasing in the longitudinal direction (along $x$-axis) to the right off the interface, starting from a finite value $\rho_{00}$ at the leftmost point of the plasma cloud. It is also supposed that the density does not change in the transverse direction. The interface separates the surrounding gas with the homogeneous distribution of the parameters $\rho_1, T_1$ and the plasma with the gas density distribution $\rho_2(x) = \rho_{00} \exp(-x/z_0)$ and the temperature $T_2(x)$. The coordinate $x$ is counted from the leftmost point on the cloud boundary and $z_0$ is the distribution characteristic length. The initially spherical shock front is incident on the spherical/cylindrical interface with the horizontal constant speed $V_1$, from left to right, center to center, and after the refraction continues to propagate in the non-uniform medium. This problem for the refracted shock parameters has been already solved in [22] using 2D model and the time dependent system of equations for the front's surface coordinates can be borrowed from there:

$$X_i = \sigma \left(t - t_{0i} + t_\lambda\right)^{2/5} + \varepsilon \left(t - t_{0i} + t_\lambda\right)^{4/5} - x_0$$

$$Y_i = y_i - \sigma \left(t - t_{0i} + t_\gamma\right)^{2/5} - \sigma \, t_\gamma^{2/5}$$

$$x_0 = \sigma t_\lambda^{2/5} + \varepsilon t_\lambda^{4/5}, \quad t_{0i} = \frac{x_i + x_b}{V_1}, \quad t_\gamma = \left(\frac{5V_1 \sqrt{T_2/T_1} \cos\alpha_i \cdot \sin\gamma}{2\sigma \cos(\alpha_i - \gamma)}\right) \tag{13}$$

where the time $t_\lambda$ is found from the solution of the following equation:

$$\frac{5}{2} \frac{V_1 \cos\alpha_i \cos\beta_i}{\sigma \cos(\alpha - \gamma)} \sqrt{\frac{T_2}{T_1}} = t_\lambda^{-3/5} + \frac{2\varepsilon}{\sigma} t_\lambda^{-1/5} \tag{14}$$

The shock velocity components can be found as

$$V_{2x} = \frac{2}{5}(\sigma \left(t-t_{0i}+t_\lambda\right)^{-3/5} + 2\varepsilon \left(t-t_{0i}+t_\lambda\right)^{-1/5}) \quad (15)$$

$$V_{2y} = \frac{2}{5}\sigma \left(t-t_{0i}+t_\gamma\right)^{-3/5}$$

The parameters σ and ε in (13-15) are related to the effective explosion energy $E$ in the thermal spot as $\sigma = \xi(E/\rho_{00})^{1/5}$, $\varepsilon = (K/z_0)\sigma^2$, $\rho_{00}$ is the density on the right side of the interface, and $\xi = 1.075$ and $K = 0.185$ are the numerical parameters borrowed from Ref. [39].

The numerical results shown in the Figs. 6-8 were produced for a relatively slow change in the distribution with $z_0 = 2.25$ cm, for $R_b = R_s = 0.1$ cm, $M_1 = 1.9$, $T_1 = 293$ K, $\rho_1/\rho_{00} = 10.0$. The gas at both sides of the interface is assumed to be ideal with the adiabatic constant $k = 1.4$, and the interface is taken as smooth [35]. The parameters $\alpha = 7$ and $\beta = 402.79$ used in the simulation correspond to the specific explosion energy $E/c_{00} = 11.707 \cdot 10^3$ J m³/kg. The dynamics of the shock front and vorticity development has been studied at several interaction times, between $n = 0.05$ and 0.40 through the equal time intervals of $\Delta n = 0.05$, and the last two times correspond to $n = 0.43$ and 0.45. The part of the shock front that is outside of the plasma cloud keeps the same spherical shape during all the interaction time (not shown in the picture).

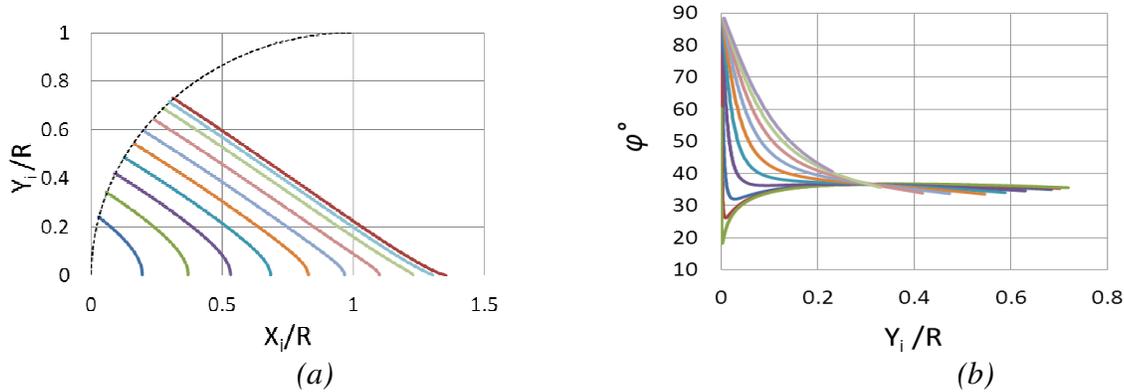

(a)                                    (b)

Fig. 6. (a) Distorted shock fronts during propagation through plasma with the exponential density distribution, for interaction times in the interval n = 0.05-0.40 with Δn = 0.05 increments, and for n equal to 0.43 and 0.45 for the two last curves. The curve sequence is from left to right. (b) The front surface inclination angle φ vs vertical coordinate Yi/R corresponding to the curves in the graph (a). The curve sequence is from upper to lower.

The specific consequence of the exponential density distribution in this problem shows up in the distinct shape of the shock front (Fig.6). Initially it appears curved close to the spherical (first few curves) and later it transforms into practically perfect cone that keeps its shape during all the time of the shock's motion through plasma (several last curves). Compared to the previous case of homogeneous distribution, in this case there are practically no areas on the shock front where its shape would change enough sharply, except a very narrow region in the close proximity to the symmetry axis. Therefore, the vorticity does not develop in most of the plasma volume except in the narrow region corresponding to the very tips of the fronts that still have sharp changes in their shapes, as shown in Fig.7.

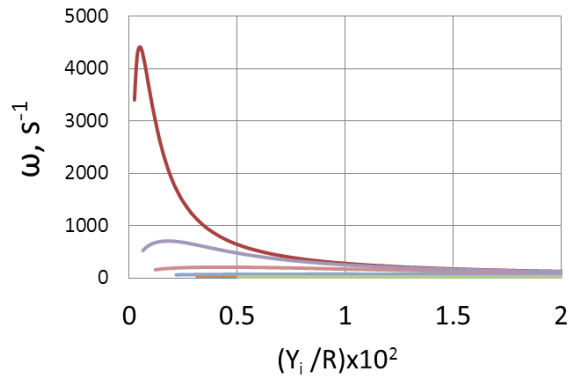

*Fig. 7. The vorticity generated in plasma with the exponential density distribution, for the same parameters as in Fig. 6. Note that due to a small size of the structure, the Y-coordinate was scaled with the factor of $10^2$ and thus the picture is greatly zoomed in the narrow region next to the symmetry axis. The curve sequence is from lower to upper.*

If scaled with the factor $V_1/R_s$, the vorticity turns out to be about two orders of magnitude less intense compared to the levels found in the uniform density distribution case. Considerably lower vortex intensity and a very small size of the structure found here can probably explain why sometimes the vortex system "does not develop" in experiments even though noticeable changes in the shock structure and plasma parameters are still present: it can simply be of non-observable size and intensity or a particular plasma parameter distribution does not allow *sharp* modifications in the front shape so no vorticity is generated.

Regardless of the smaller size of the effect, the main features of the vortex generation and their relationship with the co-processes in the flow are still preserved for this density distribution. Results for the up-stream flow rotation measured with the angle $g$ and its distribution presented in Fig. 8 are in a very good agreement with the vortex system size and its origin location shown in the Figs.6 and 7. As seen in the graphs, the flow rotation develops very sharply but in the same narrow area next to the symmetry axis where the maximum of vorticity is observed (Fig. 7).

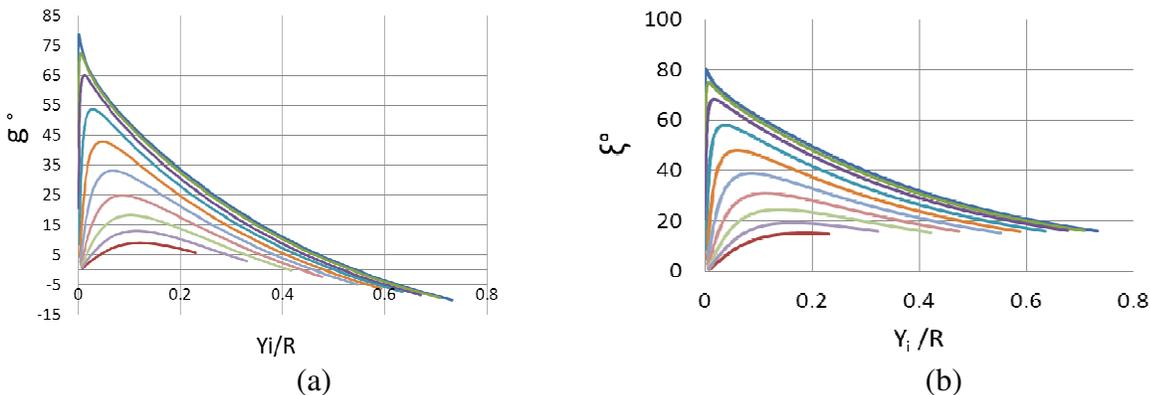

*Fig. 8. (a) The up-stream flow rotation angle g vs the vertical coordinate $Y_i/R$ for the case of exponential density distribution law, for the same parameters as in Fig. 7. (b) The flow divergence angle ξ vs $Y_i/R$. The curves sequence on both graphs is from lower to upper.*

The flow divergence angle $\xi$ is shown in the Fig. 8b and represents the amount of up-stream flow rotation relative to this for the shock velocity vector. Remarkably, the maximum vorticity intensity locations in Figs. 6, 7 are still the same as for the flow rotation angles in the Fig. 8a.

Similar to the uniform case, the pressure in the flow behind the shock and the wave drag also drop [22], confirming its direct connection to the vorticity.

Thus the results of this section represent more evidence that the shock wave refraction can trigger changes in the flow resulting in a vortex structure development. The results also show that the plasma parameter distribution may determine the intensity, the origin location, and the size of the developing vortex ring. For the exponential density distribution and the problem geometry, the vortex system originates at the tip of the almost conical shock front. As the front distortion spreads further from the tip, the system develops into a larger structure that involves more and more matter into the rotation. The vortex intensity also increases with time at all locations in the structure and the vortex centers defined by a maximum intensity tend to shift toward each other. It follows from the results that, in general, a vortex structure in an arbitrary plasma environment may not necessary originate exactly at the front tip. Sharp enough distortion at any location on the front is the condition necessary for the vortex formation. Its positive dynamics requires continuing and increasing distortion of this portion of the front with time. It follows from the results that the size of the sharply distorted shock surface area determines the size of the generated vortex structure.

### III. C. Power-law density distribution case

When the density is decreasing according to the power law, the result of the shock-plasma interaction is the subject to distinctive effects [22]. When a planar shock wave propagates through a gas with the density that drops to zero over a distance $a$, according to a power law $\rho \approx x^N$, the so-called energy cumulation effect takes place [40-42]. In the gas-dynamical approximation, a strong planar shock wave propagating in such a medium accelerates very quickly accumulating virtually infinite energy. Such a density distribution can be established in the heated spherical shells in thermodynamic equilibrium in the presence of radiative heat conduction.

Propagation of a shock through plasma with such a density distribution in the sphere-to-sphere geometry has been studied in [22] and the derivations given below will be done for the same problem geometry and parameters used in the paper. The density is assumed as changing in the longitudinal direction only starting from the leftmost point on the interface. Then the coordinates of the shock front in plasma at a point of interaction $i$

$$X_{2i} = a_i - \frac{G_i}{b}(t_{0i} - t)^b, \quad a_i = a - R_b(1 - \cos\alpha_i) \tag{16}$$

$$Y_{2i} = y_i - (V_2/V_1)\sin\gamma(nR - (x_i + x_b))$$

and the shock front velocity components can be determined as

$$V_{2xi} = G_i(t_{0i} - t)^{b-1}, \quad V_{2yi} = V_2\sin\gamma_i, \quad V_2/V_1 = \sqrt{\cos^2\alpha_i(T_2/T_1)(M_{2n}/M_{1n})^2 + \sin^2\alpha_i} \tag{17}$$

where

$$G_i = \frac{V_1\cos\gamma_i\sqrt{\cos^2\alpha_i \cdot (T_2/T_1)(M_{2n}/M_{1n})^2 + \sin^2\alpha}}{[t_{0i} - ((x_i + x_b)/V_1)]^{b-1}} \tag{18}$$

$$t_{0i} = \frac{b(a - R_b(1 - \cos\alpha_i))}{V_1 \cos\gamma_i \sqrt{\cos^2\alpha_i \cdot (T_2/T_1)(M_{2n}/M_{1n})^2 + \sin^2\alpha}} + \frac{(x_i + x_b)}{V_1} \qquad (19)$$

Here $a$ and $a_i$ are the distances to the zero density plane, counted in the direction along the symmetry axis and from the boundary at a particular point $i$ correspondingly. The time $t_{0i}$ is the local time of a point $i$ trajectory to the zero density plane, the constant $b = 0.59$ was determined in [40], and the constant $N = 3.25$ is taken from [42]. The interaction time $t$ is counted from the moment when the shock and the interface first meet each other.

This system of equations (16-19), together with the equations (2-10) have been used to simulate the data for the vorticity generation. Figs. 9-10 illustrate the front shapes and the dynamics of the vorticity development at several times of interaction, starting at $n = 0.025$ through the equal time intervals $\Delta n = 0.025$. The data was generated for the case of $R_s = R_b = 0.1$ cm, $M_1 = 1.9$, $T_1/T_2 = 0.10$, the distance $a = 3.0R_b$, and the interface was taken as smooth. Note that the interaction times used for this case are twice as short compared to the two previous cases due to considerably higher shock velocities.

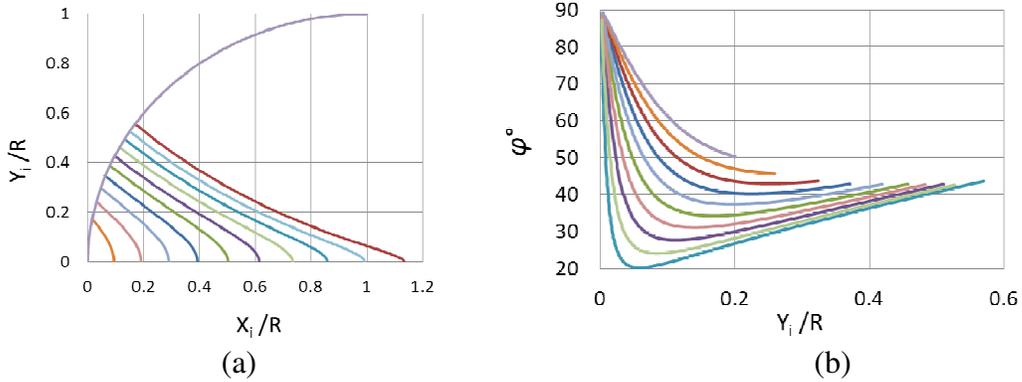

(a) (b)

*Fig. 9. (a) The shock front distortion for the case of power law density distribution, for the interaction times n = 0.025-0.250 with 0.025 increments. (b) The front inclination angle φ vs the coordinate Yi/R. The curve sequence is from upper to lower.*

The most remarkable consequence of the power-law distribution is seen in more shock front dilatation per unit of time (Fig. 9a) resulting in stronger front distortion over its entire surface. At the same time, the front inclination angle φ distribution is softer (Fig. 9b). These gentler front deformations result in a vorticity that is developing slower (Fig. 10) but still, its intensity steadily grows with time at all locations on the front. The maxima of vorticity (centers) also shift toward each other as the shock advances through the volume, following the same trend for the sharpest bending on the front to move closer to the axis. Regardless of more stretched fronts overall, the maximum vorticity intensity stays approximately on the same level as in the uniform case for the same propagation times, and this corresponds to being approximately the same as the most intense front distortions near the axis. As to the total vorticity intensity integrated over the whole plasma volume, it becomes considerably higher if to account for the contribution from the regions located further from the axis, compared to other cases of density distribution.

Thus this density distribution results in a larger vortex structure with more evenly distributed intensity in its volume.

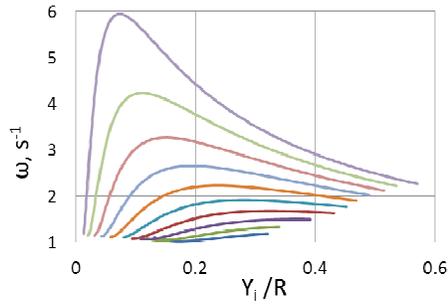

Fig. *10. The vorticity ω vs the coordinate Yi/R for the case of a power law parameter distribution, for the same interaction times as in the previous figure. The curve sequence is from lower to upper.*

The distribution of the flow rotation angle *g* versus *Yi* is shown in the Fig. 11. Compared to the uniform and exponential distribution cases, the up-stream flow velocity rotation is rather moderate here. At the same time, the curve's maximums are located further from the symmetry axis thus confirming a larger size of the vortex structure. The dynamics in the curve behavior point at the increase of the vortex ring size though its centers slowly move toward each other, with their intensity gradually increasing. Compared to two previous cases of density distribution, in general stronger and more evenly distributed front deformations result in strong vorticity that is distributed over a considerably larger volume (Fig. 10). Thus the overall size of this vortex system is larger and the structure contains a considerably larger volume of plasma matter involved in the rotational motion. It originates further from the symmetry axis, approximately at the one-quarter of the sphere radius, and then its two maximum points slowly move toward each other but never come as close as in the previous case.

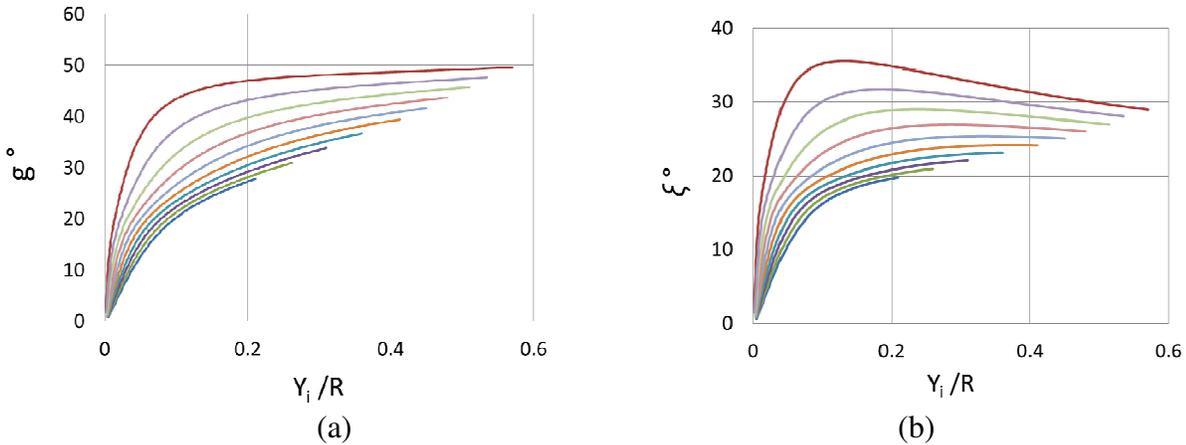

(a)                                           (b)

*Fig.11. (a) The flow velocity rotation angle g vs the coordinate Yi/R, for the same parameters as in the previous figure. (b) The divergence angle ξ vs Yi/R. The curve sequence is from lower to upper.*

Another distinct feature common for the power-law distribution is associated with the specific motion of the shock through such a medium that makes the shock stop at the plane of the zero density [40,42]. Thus, while the shock can develop virtually infinite speed, the process of the vortex development has a finite lifetime. Similar to the uniform and exponential density distribution cases, the pressure in the flow behind the shock and the wave drag drop again [22], so the relationship between the vorticity and these processes is still present.

The results found here are close to those observed in many experiments, for example [8, 12] where vortices of similar size, rotational direction, and topology were generated. The non-linear dynamics in the vorticity development closely matches the observations and evolves in the same sequence and within the same time frames with other co-processes involved in the interaction.

## IV. CONCLUSION

The results presented above showed that the shock wave refraction on an interface alone can trigger a vorticity development in the plasma volume. The key condition for a significant vortex generation is a sharp deformation of the refracted shock front followed with the up-stream flow velocity vector rotation by the angle $g$ that gives rise to a significant rotational component of the flow. For the vortex size to be observable, the deformation should span a considerable portion of the shock front, and a high enough sharpness in the front deformation will ensure significant vortex intensity. The results found here confirm many experimental observations matching the vortex topology and dynamics in its development that occur within the same time frames. The most important finding is that the proposed model explains the relationship and sequence between the generated vorticity and other processes accompanying the shock-plasma interaction. Simultaneous with the vortex generation, the pressure drop in the flow behind the shock and the following wave drag reduction in all three cases, point to the common origin and the correlation to vorticity.

It was found that while the conditions on the interface are necessary to generate vorticity, the density distribution affects the front distortion sharpness and the size of the affected area on the front and thus is responsible for such vortex parameters as its size, overall and maximum intensity, the intensity distribution over its volume, location of the vortex centers, and the motion of its centers relative to each other. The remarkably nonlinear vortex intensity growth with time found in this investigation may conceptually distinguish this mechanism from the RMI where the instability is known to initially grow with time linearly. At the same time, substantial positive dynamics in its development can separate this mechanism from baroclinicity for which the vortex intensity tends to decrease with time [7].

The vortex system with the highest intensity and of a remarkable size was produced in the plasma with the homogeneous and power law density distributions. In the latter case, the distribution of the vortex intensity was found to be less sharp compared to the case of homogeneous distribution. Overall, plasma with the power law density distribution produced larger but softer distortions on the front that resulted in a larger size of the vortex ring, softer distributions in the vortex parameters, and steady but slower dynamics in its development.

Vortex rings with remarkably distinct features were produced in the plasma with the exponential density distribution. In this case, the shock fronts developed into almost perfect cones with the surface having sharp distortions only in a very narrow area at the tip of the cone. This resulted in vorticity intensity almost two orders of magnitude lower and much smaller plasma volume affected by the rotational motion. Remarkably, the vortex parameter distributions and its dynamics were very similar to those for the two other cases of density distribution.

The cause-and-consequence relationship between the vortex formation and other co-process accompanying the shock-plasma interaction follows from the model assumptions. They can be viewed as a number of consecutive inter-related processes that occur in the following timely order. The shock refraction occurs at the moment of crossing the interface and results in the shock front distortion continuously changing with time; the distortion degree is determined by both, the conditions on the interface, and by the plasma parameter distribution; the overall distortion of the front appears as the front stretching with the degree of its distortion greatly varying along the front; this results in the redistribution of the plasma parameters behind the front; the gas pressure drop in the up-stream flow is proportional to the shock front stretching and is associated with the gradual weakening of the shock; significant drag reduction may be observed as a consequence of this [22]; if the front distortion is sharp enough, at least locally, it gives rise to a considerable up-stream flow rotation at the angle $g$ (double refraction) and the amount of rotation varies from point to point on the front; the specific distribution of this rotation in the flow results in a vortex system in the form of a toroidal ring that can grow with time if the front distortion continues to increase. The sharpness of the front distortion and the size of the front portion spanned by the distortion determine the vortex's size, its intensity, and dynamics. Thus during all the transformations, a part of the shock energy will be converted into the vortex energy that can be spent later on admixing the surrounding gas into the plasma flow.

It follows from this study that the pressure redistribution in the flow behind the shock front resulting in its significant drop can be the cause for the vorticity development rather than its consequence, as it sometimes thought in interpretations of experimental results. In fact, the vortex development can rather cause the pressure increase because of admixing of surrounding gas into the area with reduced pressure (suction effect) thus quenching the pressure lowering effect. If pressure lowering is thought, as for example in the drag reduction experiments, the vortex generation can be considered as a parasitic effect. In this case, the parameters of the shock-plasma interaction must be chosen in such a way that the shock front would be stretched considerably and as possible evenly, without sharp bendings causing the vorticity. As shown in this study, the spherical geometry and the exponential type of the plasma parameter distribution would be the best fit for this purpose. Contrarily, for combustion applications where vorticity is beneficial for better mixing in the flow, sharp distortions spanning a larger portion of the shock front are necessary and the uniform and the power law density distributions could be more desirable in this case.

The factors influencing the vorticity generation and its dynamics identified in this work can be used to control those processes in a number of applications. For example in combustion, the burning speed is controlled by introducing turbulence in the flow [3,4]. A supersonic combustion in a scramjet may also benefit from vorticity in the flow as the fuel-oxidants interface is enhanced by the breakup of the fuel into finer droplets. In studies of deflagration to detonation transition (DDT) processes show that introduction of vorticity can result in detonation. The findings of this work then suggest introducing small perturbations, for example, of the interface surface shape that will cause sharp distortions on the shock front via the refraction effect. This will result in the system of vortices and subsequent turbulence necessary for better mixing and increase in burning velocity.

In other cases though, the vorticity development must be avoid, as for example in magnetized target fusion experiments where, during the implosion of an inertial confinement fusion target, the hot shell materials surrounding the cold D-T fuel layer is shock accelerated [10]. Mixing of the shell material and fuel is not desired in this case and efforts should be made to minimize any tiny imperfections or irregularities on the front. Considering a sharp type of the interface can also be helpful since the refraction effect is up to 40% weaker in this case [35]. Elimination of vorticity in other areas of study could possibly result in reduced sonic boom, wave drag, surface temperature, and increased stability of otherwise unstable vehicle design [5].

**References:**

1. G. A. Batley, A. C. McIntosh, and J. Brindley, *The baroclinic Effect in Combustion*, Mathl. Comput. Modelling, Vol. 24, No. 8, pp. 165-176, 1996.
2. Georgievskii P. Y., *Transition to Irregular regimes of supersonic flows over bodies initiated by energy deposition*, 43rd AIAA Aerospace Sciences Meeting and Exhibit, 2005, Reno, NV.
3. A. Bret and C. Deutsch, *Beam-plasma electromagnetic instabilities in a smooth density gradient: Application to the fast ignition scenario*, Phys. of Plasmas 12, 2005, p. 102702.
4. Thomas, G. O., Bambrey, R. J. and Brown, C. J., *Experimental observations of turbulent combustion and transition to detonation following shock-flame interaction*, Combustion Theory and Modelling 5:573:594 (2001).
5. K. Kremeyer, *Lines of energy deposition for supersonic temperature/drag reduction and vehicle control*, AIP Conf. Proc. 997, 353 (2008); http://dx.doi.org/10.1063/1.2931905.
6. H. Deguchi, H. Yoneda, K. Kato, K. Kubota, T. Hyashi, K. Ogata, A. Ebe, K. Takenaka, and Y. Setsuhara, *Size and configurations in large-area RF plasma production with internal low-inductance antenna units*, Jpn. J. Appl. Phys. 45, 2006, pp. 8042-8045.
7. A. Sasoh, T. Ohtani, and K. Mori, *Pressure Effect in shock-wave-plasma interaction induced by a focused laser pulse*, PRL 97, 205004 (2006).
8. E. Shulein, A. Zheltovodov, E. Pimonov, and M. Loginov, *Experimental and numerical modeling of the bow shock interaction with pulse-heated air bubbles*, International J. of Aerospase Innovations, V.2, N.3, 2010, pp.165-187.
9. D. A. Frank-Kamenetskii, *Non-adiabatic pulsations in stars*, Doklady Akad. Nauk SSSR 80, 185, 1951(In Russian).
10. V. Suponitsky, S. Barsky, and A. Froese, *On the collapse of a gas cavity by an imploding molted lead shell and Ritchmer-Meshkov instability*, 2013. General Fusion Inc., 108-3680 Bonneville Place, Burnaby, BC V3N 4T5, Canada.
11. Fusion reactor and method for generating energy by fusion, Patent WO/2003/034441, World Intellectual Property Organization.
12. R. Adelgren, H. Yan, G. Elliot, and D. Knight, *Localized flow control by laser deposition applied to Edney IV shock impingement and intersecting shocks*, Ext. Abstr. for Jan. 2003 AIAA Aerosp. Sci. Meet., Reno, NV.
13. S. P. Kuo, D. Bivolaru, *Plasma effect on shock waves in supersonic flow*, Phys. of Plasmas, V. 8, N. 7, 2001, p. 3258.

14. A. Markhotok, S. Popovic, L. Vuskovic, *The boundary effects of the shock wave dispersion in discharges*, Phys. Plasmas 15, 2008, p. 032103.
15. A. Markhotok, S. Popovic, *Redirection of the Spherical Expanding Shock Wave on the Interface with Plasma*, Phys. Plasmas 21, 2014, p. 022105.
16. F. Marconi, *An investigation of tailored upstream heating for sonic boom and drag reduction*, AIAA paper, vol. 333, 1998.
17. Erdem, E., Kontis, K. & Yang, L., *Steady energy deposition at Mach 5 for drag reduction*, Shock Waves (2013) 23: 285. doi:10.1007/s00193-012-0405-8.
18. Jones R. T., *Aerodynamic design for supersonic speeds*, In Advances in Aeronautical Sciences, Ed. Von Karman, et all, 1959, pp.34-51, New York: Pergamon.
19. Jones R. T., The flying wing supersonic transport, Aeronaut. J., March 1991: 103-6.
20. M. Shneider, *Energy addition into Hypersonic Flow for drag Reduction and Steering, Atmospheric Pressure Weakly Ionized Plasmas for Energy Technologies*, Flow Control and Materials Processing, August 22-24, 2011, Princeton, NJ.
21. K. Satheesh, and G. Jagadeesh, *Effect of concentrated energy deposition on the aerodynamic drag of a blunt body in hypersonic flow*, Phys. of Fluids, 19, N. 3, 2007, p. 031701.
22. A. Markhotok, *The mechanism of wave drag reduction in the energy deposition experiments*, Phys. Plasmas 22, 063512 (2015).
23. Y. Ju, A. Shimano, O. Inoue, *Vorticity generation and flame distortion induced by shock flame interaction*, Symposium (International) on Combustion, Volume 27, Issue 1, 1998, Pages 735-741.
24. J. M. Picone and J. P. Boris, *Vorticity generation by shock propagation through bubbles in a gas*, J. Fluid Mech. 189: 23-51 (1988).
25. R. G. Abdel-Gayed, D. Bradley, and F. K. Lung, *Combustion Regimes and the Straining of Turbulent Premixed Flames*, Combust. Flame 76: 213-218 (1989).
26. K. N. C. Bray, Turbulent Reacting Flows, 1980, *Topics in Applied Physics*, vol. 44, ed. P.A. Libby and F. A. Williams, Berlin: Springer.
27. A. M. Khokhlov, , E. S. Orana, *Numerical simulation of detonation initiation in a flame brush: the role of hot spots*, Combustion and Flame, V. 119, Issue 4, December 1999, Ps. 400–416.
28. Batley G. A., McIntosh A. C., and Brindley J., *Baroclinic distortion of laminar flames*, Philosophical Transactions of the Royal Society of London, Series A: Mathematical and Physical Sciences, 452:199-221, 1996.
29. J. Kim, A. Matsuda, T. Sakai, and A. Sasoh, *Drag reduction with high-frequency repetitive side-on laser pulse energy depositions*, AIAA 2010-5104, 40th Fluid Dynamics Conference and Exibit, June 2010, Chicago, IN.
30. J. M. Picone, E. S. Oram, J. P. Boris, and T. R. Young, in Dynamics of shock waves, Explosions and Detonations, 1985, pp. 429-448.
31. J. Hass and B. Sturtevant, *Interaction of weak shock waves with cylindrical and spherical gas inhomogeneities*, J. Fluid Mech.181: 41-76 (1987).
32. D. Knight, *Survey of aerodynamic Drag reduction at high speed by energy deposition*, J. of Propulsion and Power, V. 24, No. 6, Nov-Dec 2008.
33. Y. Ogino, M. Tate, and N. Ohnishi, *Shock control with baroclinic Vortex Induced by a pulse energy deposition*, AIAA 2009-1225, 47th AIAA Aerospace Sciences Meeting, January 2009, Orlando, Florida
34. W. E. Baker, Explosions in Air, University of Texas Press, Austin, 1973.
35. A. Markhotok, S. Popovic, *Refractive phenomena in the shock wave dispersion with variable gradients*, J. of Appl. Phys. 107, 2010, p. 123302.

36. A. Markhotok, S. Popovic, *Shock wave refraction enhancing conditions on an extended interface*, Phys. Plasmas 20, issue 4, 2013.
37. A. Markhotok, *The Cumulative Energy Effect for Improved Ignition Timing*, Phys. Plasmas 22, 043506 (2015).
38. J. F. Liu, W. Yu, L. J. Quian, *Charge displacement in ultra-intense laser-plasma interaction*, Physica Scripta, Vol. 72 (2-3), 2005, pp. 243-246.
39. E. I. Andriankin, A. M. Kogan, A. S. Kompaneets, V. P. Krainov, *Propagation of a strong explosion in an inhomogeneous atmosphere*, PMTF 6, 1962, pp. 3-7 (In Russian).
40. G. M. Gandel'man and D. A. Frank-Kamenetskii, *Shock wave emergence at a stellar surface*, Soviet Physics, Doklady, V.1, 1956, pp. 223-226 (In Russian).
41. A. Sakurai, *On the problem of a shock wave arriving at the edge of gas*, Commune. Pure Apl. Math. 13, 1960, 353-370.
42. Y. B. Zeldovich, Yu. P. Raizer, *Physics of Shock waves and High-Temperature Hydrodynamic Phenomena*, Academic Press, New York and London, 1967.